\documentclass[sigconf]{acmart}

\newcommand\plotfigsize{0.7}

\usepackage{subfig} 
\usepackage{graphicx} 
\usepackage{todonotes}
\usepackage{confcopyright} 

\usepackage{amsmath}

\DeclareMathOperator*{\argmin}{arg\,min}

\AtBeginDocument{%
  \providecommand\BibTeX{{%
    \normalfont B\kern-0.5em{\scshape i\kern-0.25em b}\kern-0.8em\TeX}}}

\copyrightyear{2020} 
\acmYear{2020} 
\setcopyright{acmlicensed}\acmConference[ICSEW'20]{IEEE/ACM 42nd International Conference on Software Engineering Workshops }{May 23--29, 2020}{Seoul, Republic of Korea}
\acmBooktitle{IEEE/ACM 42nd International Conference on Software Engineering Workshops (ICSEW'20), May 23--29, 2020, Seoul, Republic of Korea}
\acmPrice{15.00}
\acmDOI{10.1145/3387940.3391470}
\acmISBN{978-1-4503-7963-2/20/05}

\begin{document}

\title{Insights on Training Neural Networks for QUBO Tasks}


\author{Thomas Gabor, Sebastian Feld, Hila Safi, Thomy Phan, Claudia Linnhoff-Popien}
\email{thomas.gabor@ifi.lmu.de}
\affiliation{%
  \institution{LMU Munich}
}

\renewcommand{\shortauthors}{Gabor et al.}

\begin{abstract}
Current hardware limitations restrict the potential when solving quadratic unconstrained binary optimization (QUBO) problems via the quantum approximate optimization algorithm (QAOA) or quantum annealing (QA). Thus, we consider training neural networks in this context. We first discuss QUBO problems that originate from translated instances of the traveling salesman problem (TSP): Analyzing this representation via autoencoders shows that there is way more information included than necessary to solve the original TSP. Then we show that neural networks can be used to solve TSP instances from both QUBO input and autoencoders' hidden state representation. We finally generalize the approach and successfully train neural networks to solve arbitrary QUBO problems, sketching means to use neuromorphic hardware as a simulator or an additional co-processor for quantum computing.
\end{abstract}

\ACMCopyrightPreprint{\textit{}\textit{}\textit{the 1st International Workshop on Quantum Software Engineering (Q-SE 2020)} at \textit{ICSE 2020}} 



\maketitle

\section{Introduction}
\label{sec:introduction}

Quadratic unconstrained binary optimization (QUBO) is a standard model for optimization problems (not only) in the quantum world as it can be used as input for algorithms like the quantum approximate optimization algorithm (QAOA)~\cite{farhi2014quantum} or quantum annealing (QA)~\cite{kadowaki1998quantum}. A QUBO instance of size $n$ is given as an $n \times n$ matrix $Q$ with $Q_{ij} \in \mathbb{R}$ for all $i,j \in \{1, ..., n\} \subseteq \mathbb{N}$. A solution to a QUBO instance $Q$ is a vector $x^* \in \{0, 1\}^n$ so that $$x^* = \argmin_{x} \;\; \sum_{i \leq j} Q_{ij}{x_i}{x_j}.$$

Note that QUBO instances can trivially be derived from instances of Ising spin glasses~\cite{mcgeoch2014adiabatic}. Translations to QUBO and/or Ising models exist for a multitude of common optimization problems~\cite{lucas2014ising,glover2018tutorial}, including many important NP-hard problems like 3-SAT~\cite{choi2010adiabatic} or scheduling problems~\cite{stollenwerk2016experiences}.

In this paper, we focus on the well-known Traveling Salesman Problem (TSP): A TSP instance for $m$ cities is given as an $m \times m$ matrix $D$ with $D_{kl} \in \mathbb{R} \cup \{+\infty\}$ for all $k,l \in \mathbb{N}, 1 \leq k \leq m, 1 \leq l \leq m$. A solution to a TSP instance $D$ is a vector $p^* \in \mathbb{N}^m$ that is a permutation of $(1, ..., m)$ and fulfills $$p^* = \argmin_p \;\; D_{p_{m}p_1} + \sum_{k=1}^{m-1} D_{p_k p_{k+1}}.$$

Despite apparent parallels in the formulation of a QUBO and TSP instances, the best translation from a TSP instance $D$ for $m$ cities produces a QUBO instance $Q$ of size $n = m^2$, resulting roughly in a $m^2 \times m^2$ QUBO matrix with $m^4$ matrix cells in total~\cite{feld2018hybrid}. This boost in size makes the QUBO translation rather inefficient for many practical applications and sometimes prohibits the resulting QUBO instances from being solved using quantum hardware at all, since current machines running QAOA or QA are severely limited in the amount of available qubits. However, since the computed QUBO instances originate from the smaller TSP instances, they clearly contain some redundant information.

In order to assess alternative approaches to using the limited quantum hardware for solving QUBO problems, we apply neural networks (NNs) in this paper. These can help to bridge the gap until sufficiently large quantum hardware becomes available, but also may provide hooks for additional analysis. From a black-box perspective, a NN solving QUBOs can be treated like quantum annealer by the calling modules. Having such a mockup helps to identify which aspects of software engineering are really quantum-specific solution and which originate from the problem definition.

 We explain the considered variants of NNs and how to apply them to work with problems formulated as QUBOs along the way. Using these NNs, we provide first empirical evidence for the following four hypotheses:

\begin{enumerate}
    \item Autoencoding QUBO instances generated from TSP instances is possible resulting in a hidden space having the size of the original TSP encoding (Fig.~1a, Sec.~\ref{sec:autoencoder}).
    \item NNs can be trained to solve QUBO instances generated from TSP instances (Fig~1b, Sec.~\ref{sec:tsp-qubo-direct}).
    \item NNs can be trained to solve the encoded hidden spaces of these QUBO instances (Fig~1c, Sec.~\ref{sec:tsp-qubo-encoded}).
    \item NNs can be trained to solve arbitrary QUBO instances (Fig.~1d, Sec.~\ref{sec:arbitrary-qubo}).
\end{enumerate}

\begin{figure*}[tb]
	\centering
	\subfloat[Autoencoder]{
		\includegraphics[height=8.5em]{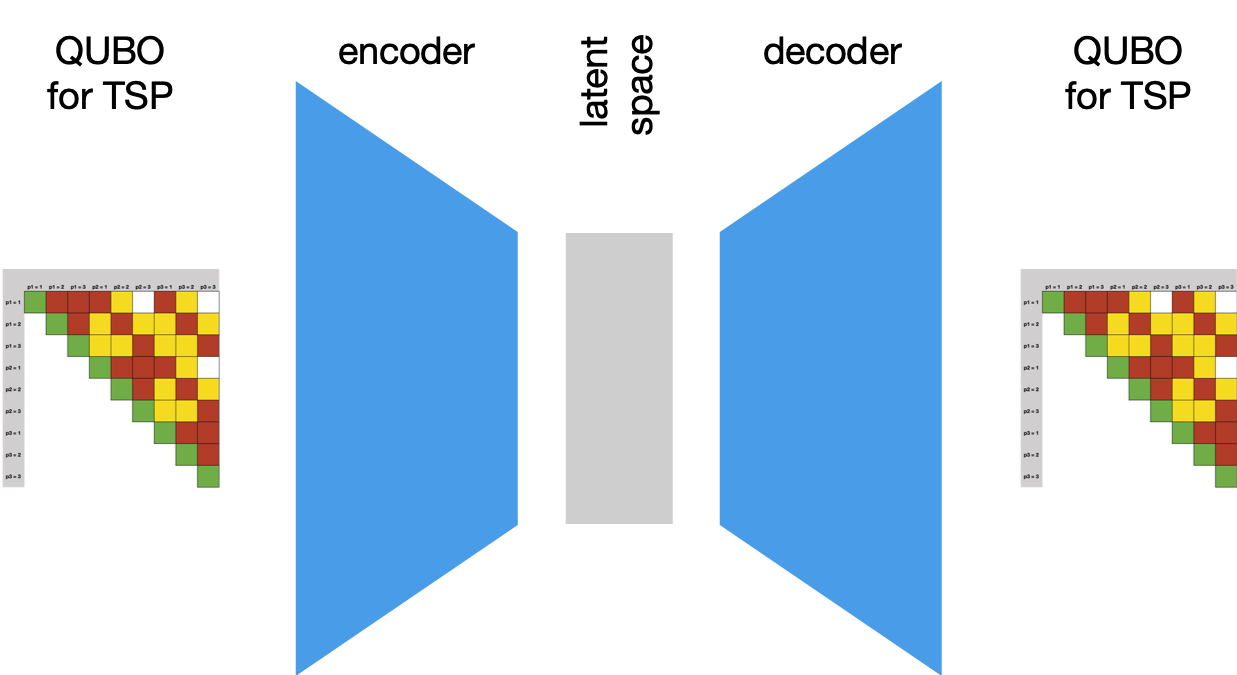}
		\label{fig:diagram-autoencoder}
	}
	\hspace{2em}
	\subfloat[TSP QUBO solver]{
		\includegraphics[height=8.5em]{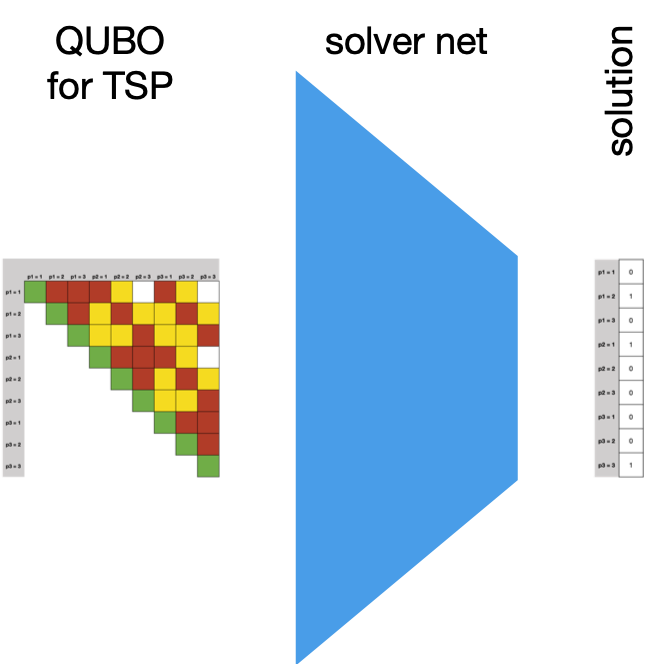}
		\label{fig:diagram-tsp-direct}
	}
    \hspace{2em}
	\subfloat[TSP QUBO encoded solver]{
		\includegraphics[height=8.5em]{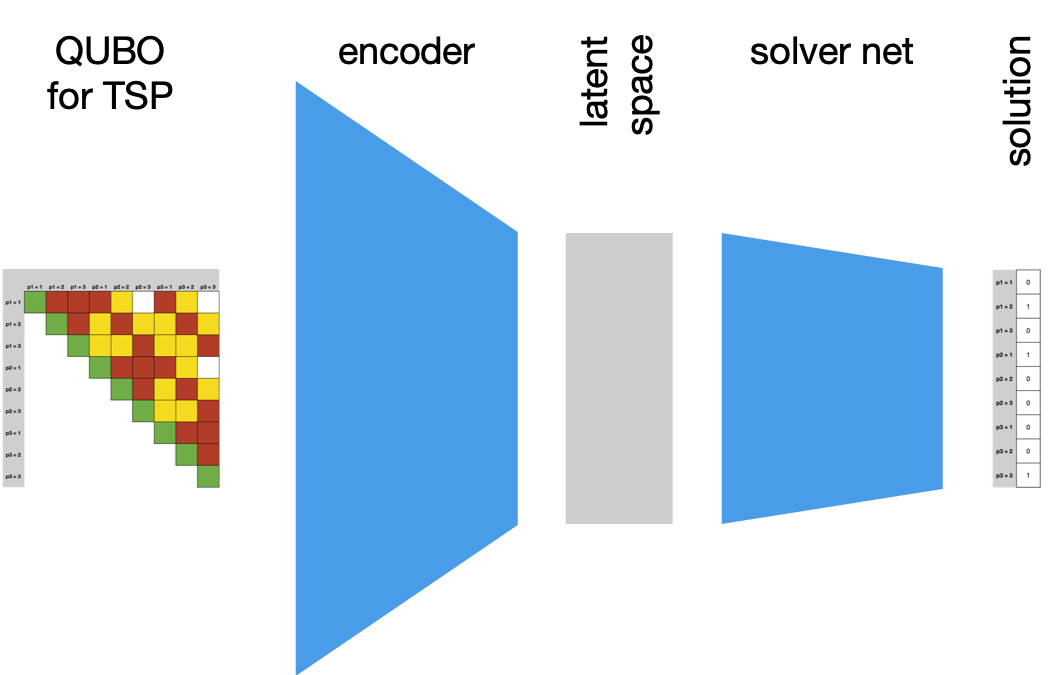}
		\label{fig:diagram-tsp-encoded}
	}
	\hspace{2em}
	\subfloat[Any-QUBO solver]{
		\includegraphics[height=8.5em]{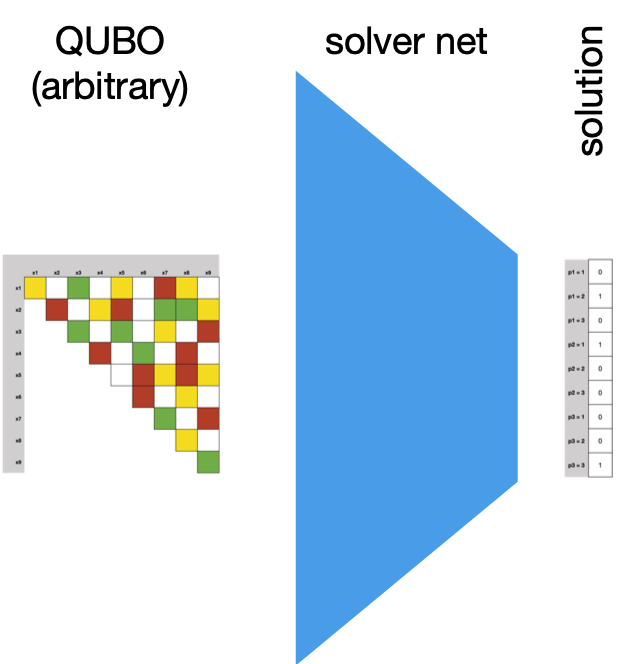}
		\label{fig:diagram-arbitrary}
	}
	\caption{Setups for testing the hypotheses.}
	\label{fig:setups}
	\vspace{-.5em}
\end{figure*}

An overview over the tested network architectures and setups is given in Figure~\ref{fig:setups}. We discuss the lessons learned from these experiments and motivate further research in Sec.~\ref{sec:conclusion}.

\section{Autoencoding QUBO formulations of TSP}
\label{sec:autoencoder}

Autoencoders (AEs) are NNs that typically possess an hourglass form: In the center they feature a hidden layer that is substantially smaller than the same-sized input and output layer. An AE is trained to reproduce its input data, but as the hidden layer is smaller than the input samples, they cannot simply ``pass through'' their inputs. Instead, the AE's first half (called \emph{encoder}) needs to learn to abstract the most relevant features so that it can populate the \emph{latent space}. This is the space of information that can be contained in the smallest hidden layer as densely as possible. Then, the second half (called \emph{decoder}) will use this representation to reconstruct the original input as closely as possible.~\cite{hub}

Once trained, AEs can be used to compress and decompress information by using the encoder and decoder part separately or to detect anomalies (i.e., input data not fitting the previously constructed latent space is assumed to substantially differ from previous training data). In our case, we use the process of training various AEs to estimate the entropy contained within the input data: the smallest latent space that still allows for almost no loss in autoencoding gives an estimate of the contained entropy in the data set, given that the encoder and decoder have been trained perfectly (if they have not, the estimate becomes rougher).

\subsection{Setup}

We have trained, tested and validated the network using different data sets. The training data consists of 11,000 randomly generated TSP instances that have been translated to QUBO; the test and validation data sets each consist of 1,000 samples.

There are different types of AEs, each with different advantages and disadvantages. The \emph{vanilla autoencoder} represents the simplest form and consists of a network with three layers. After the input layer, a dense layer with a ReLU activation function reduces the input's dimensionality, followed by a second dense layer using sigmoid as an activation function that reconstructs the input. The \emph{multilayer autoencoder} extends the previously described version by two more layers in both the encoder and decoder part. All layers use the ReLU activation function except the last layer, where the sigmoid activation function is used again. Finally, the \emph{convolutional autoencoder} uses three-dimensional vectors instead of one-dimensional vectors, which is designed to be more suitable for compressing images and tested here for compressing matrices. Our setup consists of eleven layers: starting with an input layer, there are four encoder layers, two of which are pooling layers and the other two are convolutional layers with ReLU activation function. The decoder part consists of six layers: three convolutional layers with ReLU activation function, two upsampling layers and finally an output layer that uses the sigmoid activation function.

The initial layer set for each of the AEs is inspired by \cite{hub}. Depending on the type of layer (convolutional or dense), the input data's form must be adjusted. For the convolutional layer, a QUBO matrix is represented by an array of arrays. For the dense layer, the arrays have to be flattened, so QUBO problems are represented as a one-dimensional array in order to enable the network to recognize different problems.
The final settings for each network were determined using various experiments and evaluations, which are presented in the following subsection.

The mean squared error (MSE) was used as a loss function for each AE since it shows a higher sensitivity to outliers than, for example, the absolute error. MSE calculates the average of the squared errors between predicted and actual output vectors.

The optimizers adam and stochastic gradient descent (SGD) were used to optimize the AE networks. Compared to SGD, adam, which was specially developed for training NNs, has the advantage that its learning rates are adaptive and potentially specific for each parameter. While adam uses little memory and converges faster, SGD is usually better at generalizing \cite{adab}.

We measured accuracy using two methods: The \emph{default accuracy} compares each predicted output with the actual output and returns the percentage of correctly predicted outputs. This process is repeated after each episode, with one episode corresponding to a training session on the entire input data set.
However, this accuracy is of limited interest for our motivation, since we are rather interested in whether the shortest path is returned after encoding and decoding the QUBO matrix. Therefore, the \emph{after-evaluation accuracy} was also used for training, test and evaluation. This consideration is necessary because there are at least two shortest tours in an undirected graph as for each tour there exist an opposite tour of the same length. Thus, the second accuracy uses the energy values of the solved QUBO problems, both regarding the actual qubit configuration and with the predicted ones. Accuracy is then calculated from the relationship between corresponding and all energies.

Each AE was trained for $600$ epochs.
Various learning rates were tries for the SGD optimizer, starting with a learning rate of $0.0001$, $5,000$ decay steps and a decay rate of $0.96$. There were two further training setups with an initial training rate of $0.001$ and $0.01$, respectively~\cite{mac}. For each network type, the best results were achieved using a learning rate of $0.001$.
Adam optimizer was configured with no initial learning rate and in the event of poor optimization, the mentioned configurations for learning rate and decay were set. Batch size was set to $128$.
It turned out the AE using adam optimizer showed better results than the one using SGD.

\subsection{Evaluation}

Evaluating the AEs should identify whether (and to what extent) the QUBO representation of TSP instances can be reduced while at the same time being able to reconstruct the input. For this purpose, the NNs were trained and evaluated differently, starting with no reduction in dimensionality to a reduction of one fourth of the original size.
The experiments started with TSP instances with $4$ cities ($4$-TSP), i.e., a $16\times16$-sized QUBO matrix. Even though this problem size is not challenging for computers or humans, it served as a baseline for determining the best solution.

The vanilla AE has reconstructed the QUBO well up to a size of $50\%$. After that, the (after-evaluation) accuracy was below $40\%$. The accuracy of the multilayered autoencoder (MLAE) and the convolutional autoencoder (CAE) was at least $90\%$, even with a reduction to a quarter of the original size. For this reason, the vanilla AE was not evaluated further.

When encoding TSP instances with $8$ cities ($8$-TSP), both AEs performed well; the CAE was slightly better. The after-evaluation accuracy of the MLAE is $0.95$ for $4$-TSP and $0.92$ for $8$-TSP. The CAE achieves an accuracy of $0.98$ ($4$-TSP) and $0.95$ ($8$-TSP).
The default accuracy was $0.85$ (MLAE) and $0.875$ (CAE). The average energy difference of predictions that did not correspond to the actual energy was $5.2$ for MLAE and $2.0$ for CAE.
Since CAE was best able to reconstruct the input, MLAE will not be further evaluated.

As CAE in combination with adam as the optimization function achieved the best results, this setup was chosen for the following experiments involving an encoder part.

\begin{figure}[tb]
	\centering
	\subfloat[Multilayer]{
		\includegraphics[width=\plotfigsize\linewidth]{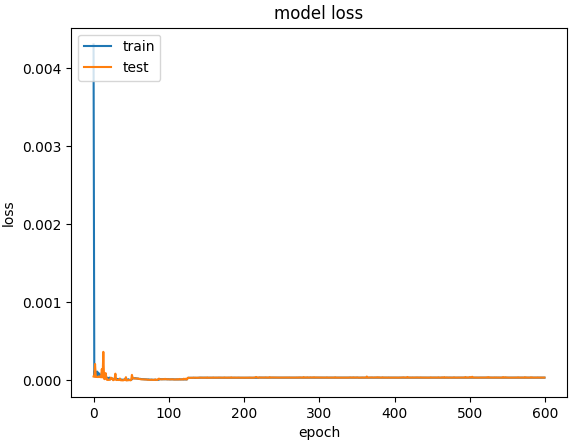}
		\label{fig:pic20}
	}\\
	\subfloat[Convolutional]{
		\includegraphics[width=\plotfigsize\linewidth]{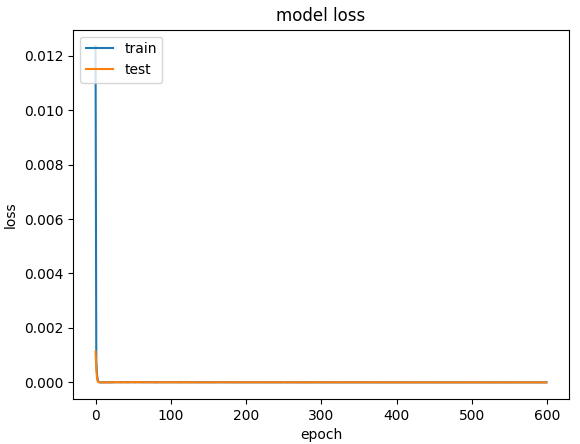}
		\label{fig:pic21}
	}
	\caption{Autoencoder model loss for $4$-TSP.}
	\label{fig:ae}
	\vspace{-1.5em}
\end{figure}

In summary, it can be said that it is indeed possible to reduce the dimensionality of TSP instances represented as QUBO problems. A reduction to a size of one fourth shows that the QUBO matrices contain lots of redundant information. If a network for outputting the correct qubit configuration can be trained just using reduced input, training time can be drastically reduced. Fig. \ref{fig:pic20} and Fig. \ref{fig:pic21} show that the reduction task is quite simple for the AEs, since training converges already in early epochs.

\section{Solving QUBO formulations of TSP}
\label{sec:tsp-qubo-direct}

The next step is to check whether a NN can be trained to solve a given QUBO problem. More specifically: is it possible to learn a qubit configuration that optimally solves a given problem.

The networks were again trained with a QUBO representation of TSP instances. However, since the required output differs from that of the AE part, new output data had to be generated accordingly. The required output for the NN is the qubit configuration for the shortest tour within the TSP instance. Corresponding qubit configurations were determined using qbsolv, a tool for operating the quantum annealing hardware by D-Wave Systems \cite{noas}. Qbsolv can also be used as a classical solver for QUBO problems.
The functionality of qbsolv regarding the solution of TSP instances up to a size of $17$ cities was checked and verified by comparing the tours returned with those calculated using Google's OR-Tools \cite{noas} as well as with the solutions of the data sets by \cite{dat}.

In order to determine a suitable NN for solving TSP instances, a recurrent neural network (RNN) and a convolutional neural network (CNN) were implemented. The results of both networks were compared, whereby again all networks were trained with a data set of size 11,000, and 1,000 samples each were used for test and validation.

\subsection{Recurrent Neural Network}
\label{subsubsec:recurrent}

Our initial network model was inspired by \cite{ib}. They used one network architecture that solves both TSP and the likewise NP-complete knapsack problem. Their network uses the two-dimensional coordinates of the cities as input and the sequence of the cities to be visited as output. In our work, however, the input are TSP instances represented as QUBO matrices and the output is the shortest tour coded as a qubit configuration.

We use a pointer network consisting of two recurrent NN modules (encoder and decoder). 
As in \cite{ib}, we implement attention using long short-term memory (LSTM) cells \cite{lih}.

The loss is calculated using binary cross-entropy. This loss function is suitable for problems with a yes/no decision, which is the case with our 0/1 output representing the qubit configuration.

With regard to the optimizer function for training the RNN, we have strictly adhered to the structure of \cite{ib}. They propose to use optimization via policy gradients instead of a supervised loss function (as for the AE mentioned above). The reason for this is that the model's performance may be linked to the label's quality. 

For this, a Monte Carlo approach was implemented in the reinforcement algorithm in order to implement the policy parameters update using random sampling \cite{mad}. In addition to this model-free approach, adam was used as an optimization approach.
Again, default accuracy was used during training and subsequently after-evaluation accuracy was used for evaluating the QUBO data.

\subsection{Convolutional Neural Network}
\label{subsubsec:convolutional}

Any QUBO data can be represented as a two-dimensional matrix, which is why we also implemented a convolutional neural network (CNN). Our CNN consists of six convolutional layers and two dense layers. All but the final layer are paired with a ReLU activation. The final layer includes a softmax activation function.
The CNN's training was optimized with adam.

The first round of experiments was trained using binary cross-entropy as loss function. The network loss decreased as desired, but the accuracy did not increase. After analyzing the predicted outputs, it was found that the qubit configuration was incomplete. Most of the time, only two or three cities were visited within a TSP instance of $4$ cities ($4$-TSP), or three to four cities with a TSP instance of $8$ cities ($8$-TSP).
This observation led to changing the loss function. The binary cross-entropy function has been extended by a function that checks how many qubits are set to $1$. The function increases the loss if the number of qubits set does not match the number of cities. In addition, the loss is increased if not every city was visited, but a certain city several times.

\subsection{Setup}

The RNN TSP solver was first trained with coordinates of the cities. This was to check whether the network, which was inspired by \cite{ib}, gave similar results. It was trained on $10$-TSP and $20$-TSP and actually delivered similar results.

Then problems in QUBO representation were used as input and the resulting qubit configuration as output. Batch size was set to $128$ and the network was tested with $128$ and $256$ hidden units per layer. The range of learning rates has been as with the AE.
To save training time, the RNN was first tested with $128$ hidden units and three learning rates. The loss was best at learning rate $0.001$ with a decay of $0.96$ at $5,000$ steps. However, training with $128$ hidden units resulted in a network that was not able to recognize the hidden logic within the qubits for problems with more than $4$ cities. Accordingly, the hidden units were increased to $256$. This lengthened the training time, but the entire logic of QUBO, which represents the TSP, could still not be learned.

We suspect that the problem lies in the layers used, because -- as can also be seen with the AEs -- convolutional layers process QUBOs better. Since a further increase in the hidden dimensions would lead to a further increase in training time, we just focused on CNN for further analysis.

The CNN was trained and compared with $128$ and $256$ units per layer. Training the network containing $128$ units with $4$-TSP instances worked well, but the model overfitted. This is because the network is designed for complex problems, but a TSP with $4$ cities is just too simple. To prevent overfitting, dropout layers that randomly ignore units were added to the model when training with $4$-TSP instances.

$4$-TSP instances were used to train the $128$ units model, while $8$-TSP instances were used for models having $128$ and $256$ units (but no dropout layer).

\subsection{Evaluation}

Before the loss function was adjusted as already described, the forecast did not set $n$ qubits to $1$, but only two or three. The network afterwards learned that the goal is to minimize the energy and therefore has to consider all constraints.

\begin{figure}[tb]
	\centering
	\subfloat[$4$-TSP, $128$ units, dropout layer]{
		\includegraphics[width=\plotfigsize\linewidth]{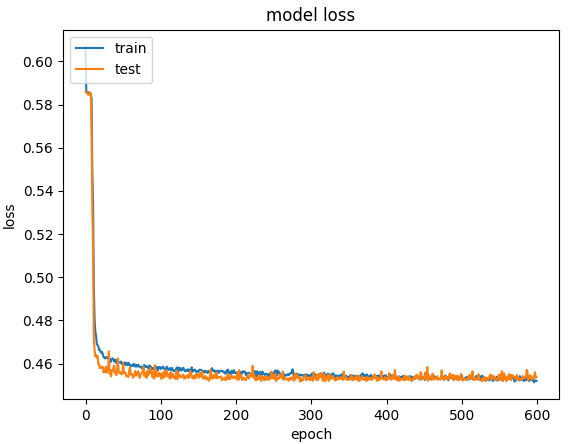}
		\label{fig:pic23}
	}\\
	\subfloat[$8$-TSP, $256$ units, pre-trained]{
		\includegraphics[width=\plotfigsize\linewidth]{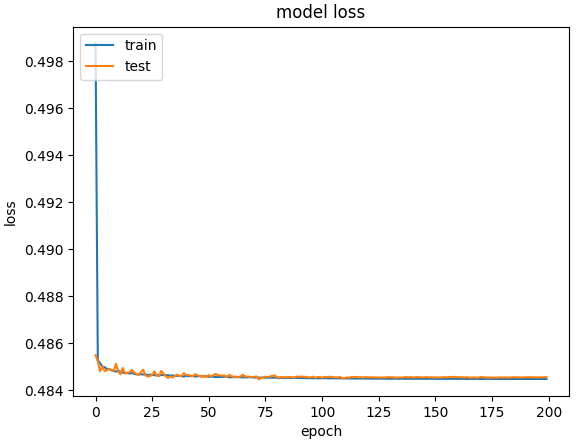}
		\label{fig:pic24}
	}
	\caption{Convolutional neural network model loss.}
	\label{fig:cnn}
	\vspace{-1.5em}
\end{figure}

The $4$-TSP setup was trained with $600$ epochs. However, the training itself only required $400$ epochs for the ideal result. After the dropout layer was added, the network no longer overfitted and showed a loss of around $0.44$ (see Fig. \ref{fig:pic23}). In $88\%$ of the cases, the predicted values matched the actual values. In cases where they did not match, the average difference between the actual and calculated distance of the shortest tour was $9.36$. If one considers that the distances were chosen randomly between $1$ and 10,000, the network did understand its task.

When training the $8$-TSP, the dropout layer was not used. $128$ units were not sufficient to achieve good results: a default accuracy of $0.14$ was achieved. After an update to $256$ hidden units and still no dropout layer, an after-evaluation accuracy of $0.65$ was achieved. The average distance for non-matching actual and predicted data was $20.15$.

Fig. \ref{fig:pic24} shows the training of $8$-TSP. One recognizes that the loss starts lower than with the $4$-TSP. A major disadvantage of convolutional neural layers is the training time. In order to save processing time, all $8$-TSP instances were trained with pre-trained networks. The pre-trained networks are networks that were trained using $4$-TSP. This procedure helps to reduce the processing time, since the loss starts at a lower point because only the last layers have to be trained. It also leads to fewer epochs for training convergence. In this specific case, $200$ epochs were sufficient.
We also checked that these results are similar to a CNN that was trained on $8$-TSP without pre-trained layers. The training took four times longer, the results were worse after $200$ epochs, but approximately the same after $600$ epochs.

\section{Solving encoded states of QUBO formulations of TSP}
\label{sec:tsp-qubo-encoded}

We now present a network architecture for solving NP-complete problems that uses the encoder part of the CAE combined with the CNN TSP solver (see Fig. \ref{fig:diagram-tsp-encoded}). The idea is to reduce the dimensionality of the QUBO problems and use this representation to train the network solving the problem. The networks mentioned were chosen because the CAE showed best results on reconstructing QUBOs and the CNN TSP solver accordingly performed best when solving TSP instances. When combining the networks, the setup as described in the previous sections was used.


Training the CNN with compressed QUBO data from $4$-TSP instances again led to overfitting. Thus, dropout layer were added to address this problem. Instances of $4$-TSP were only tested with $128$ units because the results were good enough. The training of the combinatorial NN had very similar results to the CNN. The loss converged at $0.46$ and had a default accuracy of $0.75$ and an after-evaluation accuracy of $0.85$. The average difference between all non-matching and actual results was $9.56$. The network was trained over $600$ epochs.

The compression of the input had almost no effect on the network's ability to learn qubit configurations. The network's training time was highly reduced when only compressed input was used. The CNN used about $9-10$ hours of training time, while the combinatorial NN only used about $4-5$ hours.

\begin{figure}[tb]
	\centering
	\includegraphics[width=\plotfigsize\linewidth]{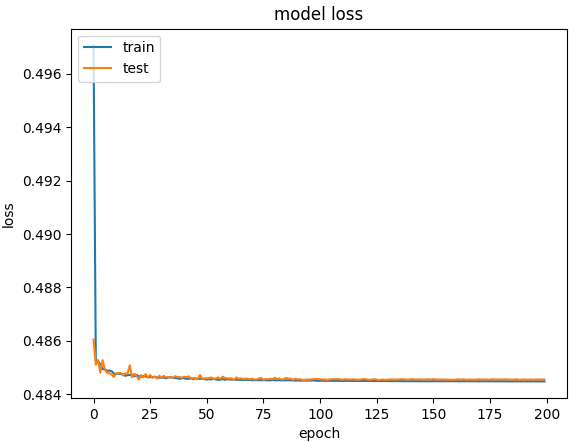}\\
	\caption[TSP 8 Combinatorial Neural Network]{Combinatorial NN model loss, $8$-TSP, $256$ units, pre-trained.}\label{fig:pic24-neu1}
	\vspace{-1em}
\end{figure}

In order to learn $8$-TSP instances, the combinatorial NN has again used pre-trained layers, i.e., those of the $4$-TSP combinatorial NN. Again, there are only minor differences from the CNN results. The loss was $0.48$ (see Fig. \ref{fig:pic24-neu1}), which is identical to the CNN's loss. The network was trained for $200$ epochs, had a default accuracy of $0.55$, an after-evaluation accuracy of $0.64$, and a mean difference between non-matching and actual results of $27.5$. This value is $7.35$ higher than that of the CNN, but still acceptable as the cities' distances were randomly chosen between $1$ and 10,000.

\section{Solving arbitrary QUBO instances}
\label{sec:arbitrary-qubo}

Finally, we want to take another step towards generalization and train NNs to solve arbitrary QUBOs. In this way, they can be functionally used in place of a quantum annealing solver.


\subsection{Setup}


Random QUBOs have no inherent structure that could be exploited by an AE, so we only trained CNNs for this task. The input data was generated by filling the upper triangular matrix with random numbers between $-$10,000 and 10,000. The output was generated by labeling given input with the qubit configurations that were created using qbsolv. The training data set consisted of 11,000 samples, the validation and test data set each consisted of 1,000 samples.

\subsection{Evaluation}

We want to show that a single NN can solve not only a specific NP-complete problem, but a generic one. In order to obtain comparability, random QUBOs were created that have the same dimensionality as $4$-TSP and $8$-TSP QUBOs.


The CNN was able to learn from the random QUBOs: $16\times16$-dimensional matrices (equivalent to $4$-TSP) were trained using $256$ units per layer over $400$ epochs and had a loss of $1.06$. The default accuracy was $0.45$, the after-evaluation accuracy $0.48$. The mean energy difference between non-matching actual and predicted results was $230.32$, which is much higher than with TSP. However, the qubit configuration was correct for almost every second result.

\begin{figure}[tb]
	\centering
	\subfloat[$16\times16$]{
		\includegraphics[width=\plotfigsize\linewidth]{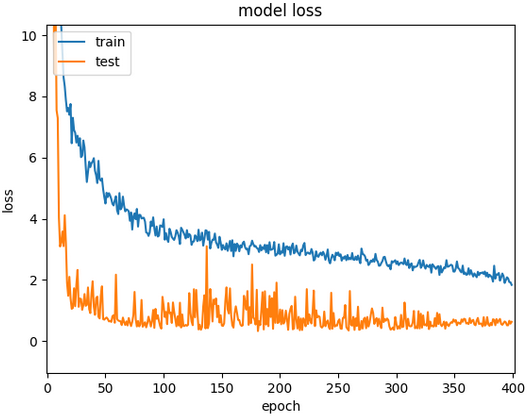}
		\label{fig:pic22-neu2}
	}\vspace{-.5em}\\
	\subfloat[$64\times64$]{
		\includegraphics[width=\plotfigsize\linewidth]{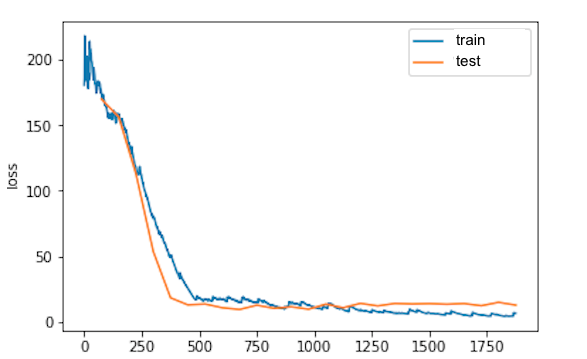}
		\label{fig:pic23-neu3}
	}
	\caption{Random QUBO Solver model loss.}
	\label{fig:random}
		\vspace{-1.5em}
\end{figure}

Training using random data is far more complex than training a specific problem (see Fig. \ref{fig:pic22-neu2} and Fig. \ref{fig:pic23-neu3}). The network did not overfit, not even with twice as many units. The random $64\times64$-sized QUBO problems (equivalent to $8$-TSP) were trained with a network having $256$ units per layer over $1,800$ epochs and had a loss of $9.05$. Default accuracy and after-evaluation accuracy were around $0.2$ and the average energy difference of the non-matching outputs was $345.45$.

Training with random values took a lot of time for relevant QUBO sizes, the accuracy fell faster with the increase in the QUBOs' dimensionality than with TSP. The use of a network pre-trained on $16\times16$-sized random QUBO problems inside a $64\times64$ random QUBO network had an accuracy of $0.12$ and did not work comparable to the CNN. In addition to the fact that an energy minimum is sought, the larger network cannot reuse much information.

\section{Conclusion}
\label{sec:conclusion}

We provided empirical evidence for four hypotheses. (1) AEs are able to filter the overhead induced by a QUBO translation of TSP to some extent. They can thus be used to guess the original complexity of a problem from its QUBO formulation. (2) NNs can be trained to return the qubit configuration resulting in minimum energy for a QUBO problem generated from a TSP instance. They are thus able to solve TSP even in a larger QUBO translation. (3) Accordingly, NNs can also solve QUBO problems originating from TSP given their latent space representation (instead of the full QUBO matrix). (4) NNs can be trained to solve QUBO problems in general. The fact that CNNs appear most effective implies that QUBO problems can be treated more like a somewhat local graph problem and less like combinatorial optimization.

These first steps call for immediate follow-up research. Most importantly, a thorough study of the various impact of overhead from the QUBO translation is necessary: How do networks that have been trained for (a) solving TSP in native encoding, (b) solving QUBO translations of TSP, and (c) solving QUBO in general compare on the same set of problems regarding various performance metrics? Are there cases where a QUBO translation may actually be easier to solve than other representations of TSP? Does specialized training on just one type of QUBO bring any advantage over training on random QUBOs? How do the results on TSP (whose QUBO translation introduces a quadratic overhead) compare to problems with more (or less) efficient QUBO translations?

From this experience report, a strong argument can be made for mathematically solid interfaces in quantum computing: The NNs we trained should be able to replace any other means of solving QUBOs fully transparent to the provider of the problem instances. A diverse pool of mechanisms for solving QUBOs should prove useful to establish QUBO as a suitable formulation for optimization problems and thus prepare for the eventual deployment of quantum-based machines. Current breakthrough technology like neuromorphic hardware may thus serve as a bridge to the quantum age.

We argue that for some time to come, quantum software will usually only be shipped as a module within larger, mostly classical software applications. Furthermore, these modules will usually come with fully classical counterparts as quantum resources will remain comparatively limited and thus should not be used up unnecessarily, for example when testing other parts of the software where a good enough approximation of the quantum module suffices. We think that NNs may provide a very generic tool to produce such counterparts as it has been done in this case study for quantum annealing or QAOA, even though their rather black-box nature opens up a new field of testing issues. Effectively, we argue that any approach to the integration of quantum modules should aim to include similar classical approximation models at least for the near future.

We would like to point out that even in the presence of large-scale quantum hardware, handling QUBO problems with NNs might still be useful for pre- and post-processing of problem instances, dispatching instances to various hardware platforms, or providing estimates of the inherent complexity of a specific problem or problem instance. As we have shown that NNs can handle the structure of QUBO matrices well, they may also be able to learn transformations (ideally with automatic reduction of size) on them or help with introspection of the optimization process and effectively the debugging of optimization problem formulations or quantum hardware platforms.

\bibliographystyle{ACM-Reference-Format}
\bibliography{main,my}

\end{document}